\documentclass[12pt]{iopart}
\pdfoutput=1
\usepackage{color}
\usepackage[pdftex]{graphicx}
\usepackage[pdftex]{hyperref}
\hypersetup{colorlinks=true,citecolor=blue,linkcolor=red,urlcolor=blue}
\usepackage[all]{hypcap}
\usepackage[normalem]{ulem}
\usepackage{etoolbox}

\expandafter\let\csname equation*\endcsname\relax

\expandafter\let\csname endequation*\endcsname\relax
\usepackage{amsmath}
\begin{document}

%opening
\title{\textbf{Modeling nanoribbon peeling}}
\author{\large{L Gigli,\textit{$^{a}$} A Vanossi\textit{$^{b,a}$} and E Tosatti\textit{$^{a,b,c}$}}}

\begin{abstract}

The lifting, peeling and exfoliation of physisorbed ribbons (or flakes) of 2D material such as graphene 
off a solid surface are common and important manoeuvres in nanoscience. The feature that makes 
this case peculiar is the structural lubricity generally realized by stiff 2D material contacts. 
We model theoretically the mechanical peeling of a nanoribbon of graphene as realized by 
the tip-forced lifting of one of its extremes off a flat crystal surface. 
The evolution of shape, energy, local curvature and body advancement are ideally  expected to follow 
a succession of regimes: (A) initial prying, (B) peeling with stretching but without sliding (stripping), 
(C) peeling with sliding, (D) liftoff. 
In the case where in addition the substrate surface corrugation is small or negligible, then (B) disappears, and we find that 
the (A)-(C) transition becomes universal, analytical and sharp, determined by the interplay between bending rigidity and 
adsorption energy. This general two-stage peeling transition is identified as a sharp crossover in published  
data of graphene nanoribbons pulled off an atomic-scale Au(111) substrate.

\end{abstract}

\maketitle

\footnotetext{\textit{$^{a}$~International School for Advanced Studies (SISSA), Via Bonomea 265, 34136 Trieste, Italy}}
\footnotetext{\textit{$^{b}$~CNR-IOM Democritos National Simulation Center, Via Bonomea 265, 34136 Trieste, Italy}}
\footnotetext{\textit{$^{c}$~The Abdus Salam International Centre for Theoretical Physics (ICTP), Strada Costiera 11, 34151 Trieste, Italy}}

\section{Introduction}
\label{sec:Intro}
2D materials, combining atomic thickness, extreme robustness and substantial flexibility, have become extremely important 
in condensed matter physics and materials science. The basic actions necessary to produce and handle them involve nanomanipulations  
such as exfoliation, deposition, dragging, lifting and peeling. A notable example which probes the contact and frictional properties of 
graphene-metal interfaces is the peeling of graphene sheets ~\cite{Miskin18} and the sliding, lifting and eventual 
detachment off metal surfaces of graphene nanoribbons (GNRs) ~\cite{Kawai16, Gigli17, Gigli18, Gigli19}.
The physics of  peeling of a 2D adsorbate off a substrate on one hand, and the sliding friction on that substrate on the other hand, are intimately related.
In the strong friction limit, the  detached (and stretched) part of length $l' + l_C$ is being stripped, as Kendall 
described long ago studying the mechanics of thin film detachment ~\cite{Kendall75}, thus consuming the adhered film part of length $l$.
In that case, after an initial prying phase which we denote by (A) (that is generally neglected), there follows a phase which we call (B),
where the peeling proceeds simply by the advancement of the detachment point between the stripped and the adhered part, whose tail end is immobile 
until liftoff, (D), where $l = 0$.
At the nanoscale, the very recent lifting and peeling of a DNA strand off an Au(111) surface appears to be qualitatively similar, owing to a substantial 
pinning of DNA by the surface corrugation ~\cite{Pawlak19}.

The situation may be quite different for other atomically thin 2D materials, and in particular it is different for graphene nanoribbons adsorbed on the same metal surface.  
In that case, weak physical adhesion prevails,  and corrugation against nanoribbon sliding is small enough to give rise to structural lubricity, 
also called superlubricity ~\cite{Hirano93, Vanossi13, Martin18, Hod18}. 
The frictional pinning of a structurally lubric film or ribbon is entirely due to 
defects. In a perfect nanoribbon deposited on a step-free surface, the only defects are the nanoribbon's perimetral edges. 
As a consequence, the overall sliding friction of a deposited nanoribbon can at most grow with a power $\alpha < 1/2$  of the 
adhered area $A = Lw$ where $L$ is the length and $w$ the width.
For Au(111)-deposited GNRs in particular the front and tail edges have been
identified as the only sources of friction, whereas the side edges are essentially frictionless.
Indeed, the GNR sliding friction on gold has recently been found to be independent of length, and therefore 
independent of adhered area ~\cite{Kawai16, Gigli17}, that is $\alpha \sim 0$.
That evidence of structural lubricity of the GNR/metal contact reflects a weak mutual corrugation (despite a large adhesion energy),
in turn associated with the GNR's incommensurability with the metal surface lattice periodicity. 

%Fig. 1
\begin{figure}
  \centering
  ~~~~~~~~~~\includegraphics[width = 0.8\columnwidth]{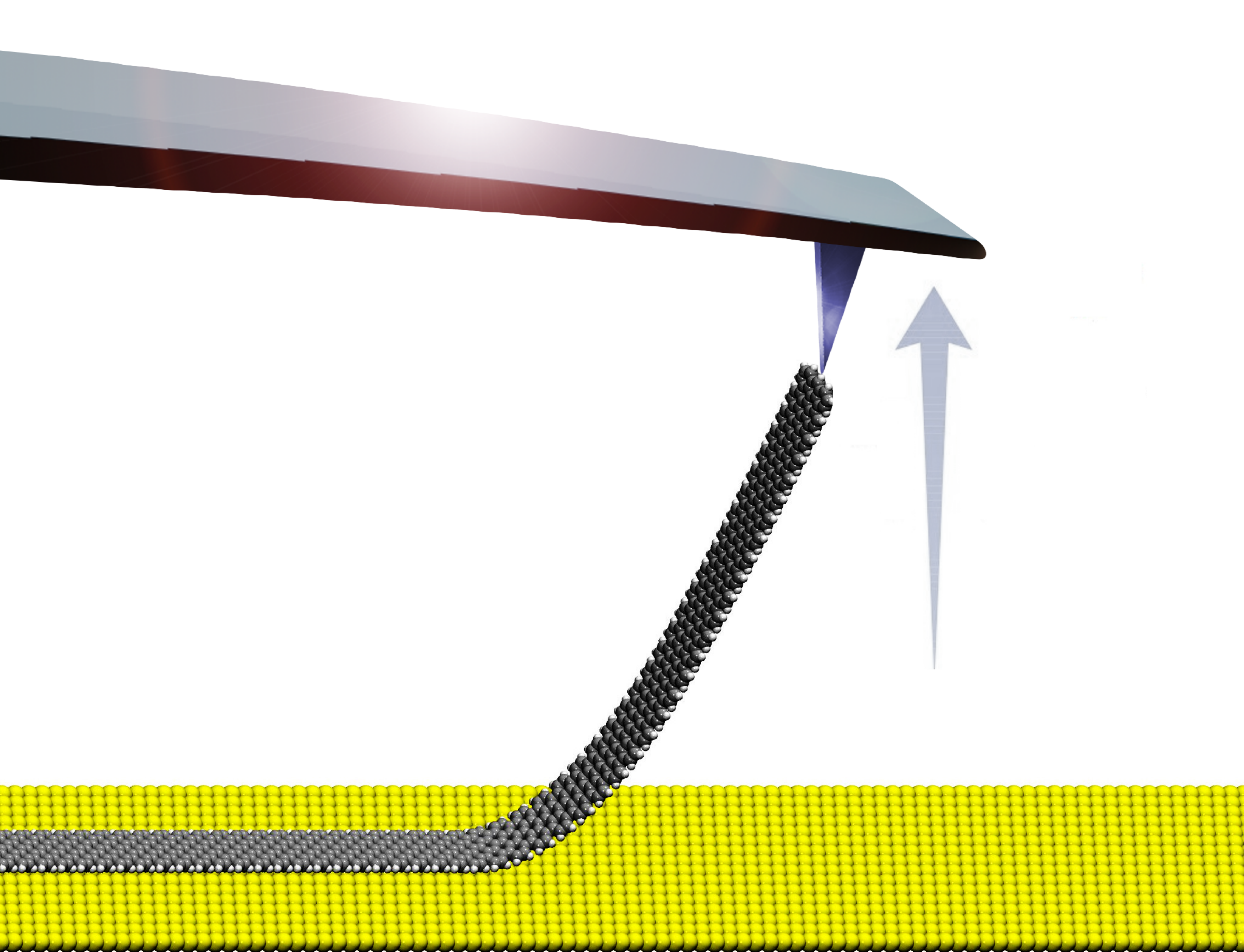}
  \caption{
    Sketch of the detachment of a several-nanometer-long armchair GNR from the Au(111) 
surface upon vertical lifting of one end by an AFM tip.
  }
  \label{fig:sketch}
\end{figure}

The mechanical lifting by one end of a GNR (or of a porphyrin derivative ~\cite{Pawlak16}) 
on a metallic surface, sketched on Fig.~\ref{fig:sketch}, recently demonstrated experimentally ~\cite{Kawai16} 
represents perhaps the simplest example of 2D nanoribbon lifting, peeling and detachment. That offers
the chance for a more general understanding of the underlying physics. 
If the interface corrugation and its effects are neglected, the peeling of the GNR by lifting of its front end 
should in general proceed through a different sequence of phases: first prying (A), then the stripping 
regime (B) should disappear together with pinning, replaced by a novel extensive regime (C),
where peeling proceeds with forward sliding of the adhered part, until liftoff (D).
In the alternative strong pinning limit described by Kendall, the sticking regime (B) prevails 
and the slipping (C) disappears.
%
%Fig. 2
\begin{figure}
  \centering
  ~~~~~~~~~~\includegraphics[width = 0.8\columnwidth]{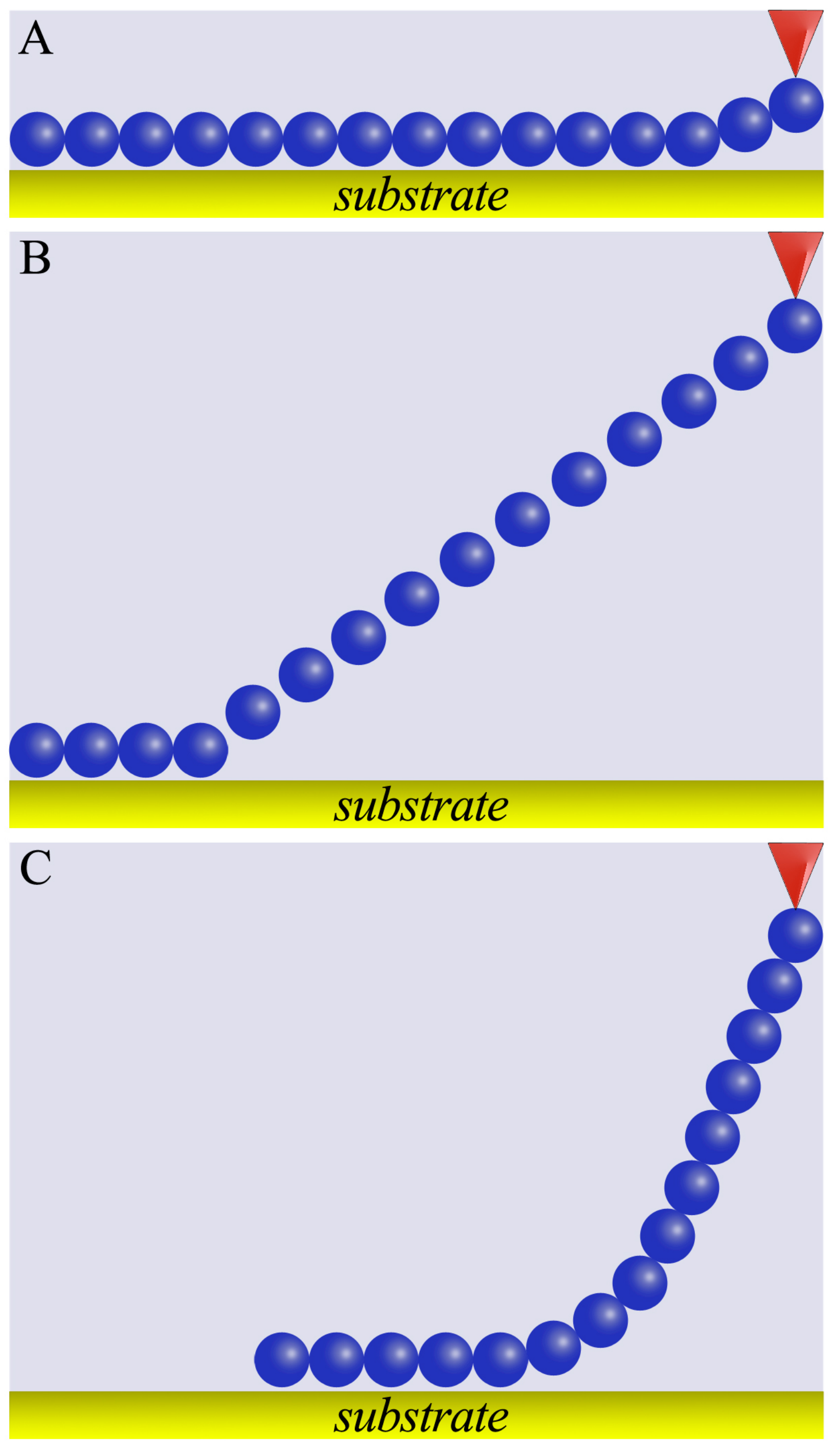}
  \caption{
    Sketch of the sequence of phases or regimes that a nanoribbon may undergo upon peeling by mechanical lifting: (A) prying; (B) stripping without sliding; (C) peeling with lubric sliding (signaled by the advancing tail end).  In the limit of zero corrugation treated in our analytical model, regime (B) disappears. Conversely, (C) will be absent  in the opposite strong pinning (Kendall) limit. Note that stretching of the detached part which we ignore in our model is important in (B) as shown,  but is small and  negligible in case (C). 
  }
  \label{fig:peeling}
\end{figure}
In the real, intermediate GNR case, the residual effect of weak corrugation can be expected to mix regimes (B) and (C) in  time. 
That is confirmed by our previous simulations ~\cite{Gigli19}, where the GNR advancement indeed takes place 
by an alternation of sticks (B) and slips (C).  
Examining, as we shall do here, the ideal limit of zero corrugation is nonetheless valuable. It permits to build and solve analytically a model
of the mechanical lifting, peeling and detachment of a narrow, inextensible and flexible 2D membrane-like material strip from a flat underlying surface. 
In that way we shall find and describe analytically a sharp first-order like transition between the (A) and (C) regimes -- a transition  
caused by the competition between the adhesive strength of the interface and the bending rigidity of the membrane.  As will be shown,
this transition or sharp crossover is clearly identifiable in published experimental data of GNR peeling \cite{Gigli18}, once marginal 
corrugation-induced wigglings are ignored.

\section{Smooth peeling model}
\label{sec:peel_model}

%Fig. 3
\begin{figure}[!t]
\centering
\includegraphics[width = 0.50\textwidth]{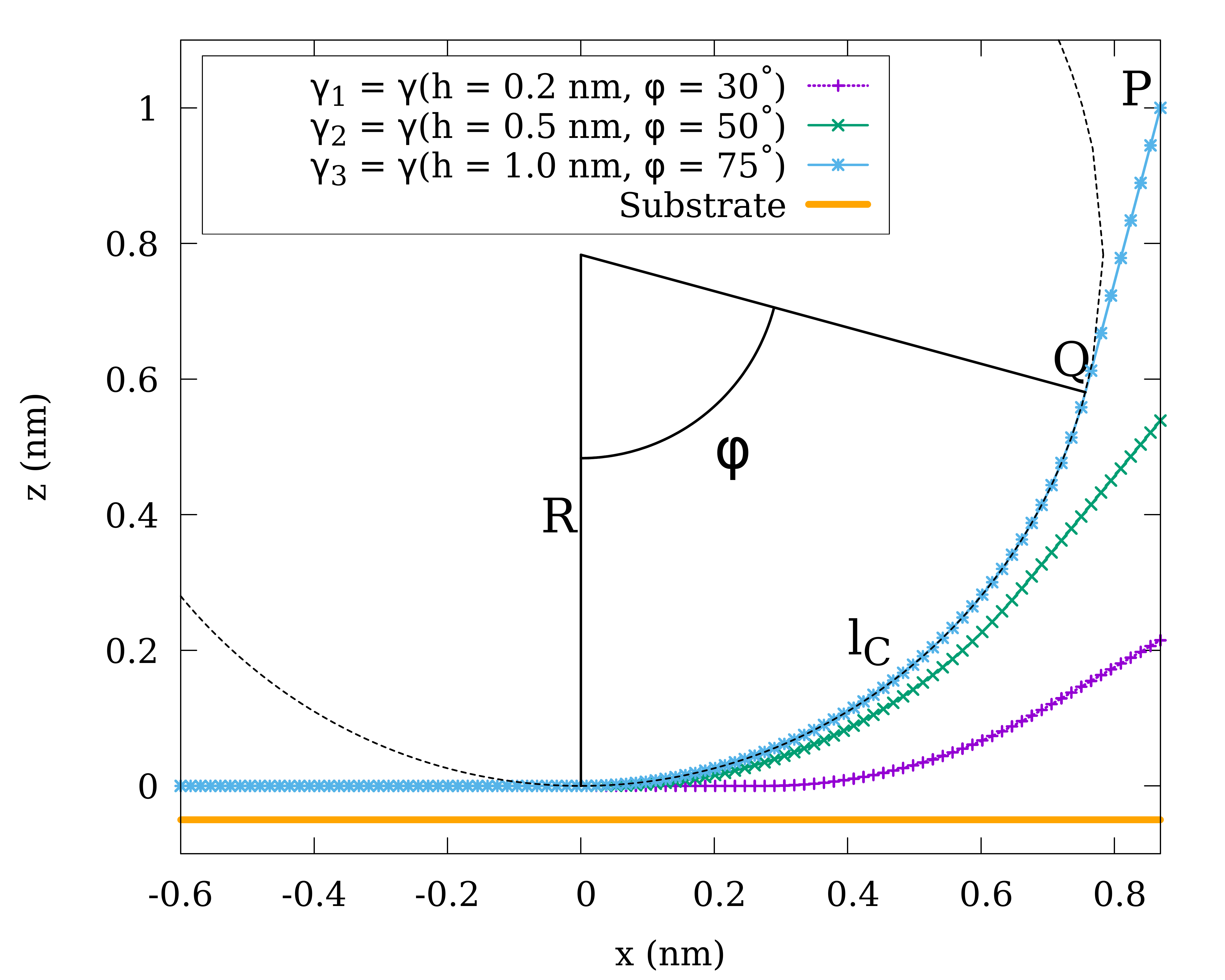}
\caption{
  Sketch of the lifted nanoribbon shape $\gamma = \gamma(h)$ for three different values of the lifting height $h$ and the corresponding 
  bending angle $\varphi$. These represent equilibrium configurations of the smooth peeling model, corresponding to solutions of Euler's 
  equations (\ref{eq:Eulero}). Parameters used, specified in text, are chosen to approximate the realistic situation of Ref. ~\cite{Gigli18}.  
  For the sake of clarity, the three curves are presented for the real situation where the $x$-coordinates of the ribbon suspension point P 
  coincide in all cases, as in experiments. Note instead that in all formulas below the $x$-reference frame is shifted for convenience so 
  as to keep the origin fixed at the detachment point. The angle formed by the detached segment QP (of length l') and the horizontal axis 
  is equal to $\varphi$, whence  $l' \sin\varphi = h – R(1 – \cos\varphi)$.
}
\label{fig:elastic_mod}
\end{figure}

The model consists of a narrow strip of length $L$, lying initially flat on a surface plane to which it adheres with an 
energy per unit area equal to $-\epsilon$. The strip, of narrow width $w \ll L$, is strong but flexible with bending rigidity 
per unit length $Kw$ (where $K$ is the usual bending rigidity). It is lifted at one extreme which is forced above the surface to 
a  height $z = h$ while its planar coordinate remains strictly fixed, as a hypothetical macroscopic pickup tip would do.
For convenience  we will use a reference frame whose origin $x = 0$ coincides with the front end contact point between 
the GNR and the flat substrate. In that reference frame the coordinates of the lifted extremity pickup point is $P = (x(h), 0, h)$, 
where $x(h) > 0$  measures the planar projection of the lifted up piece (see Fig.~\ref{fig:elastic_mod}). 
The lifting is treated as an adiabatic process at $T = 0$, beginning with $h = 0$ and ending with complete peeling when $l = 0$ and $h = L$ 
(being $l$  the length of the adsorbed fragment).
At a generic lifting $h$ the strip comprises three pieces: 
\begin{equation}
 \label{eq:lengths}
  L = l + l' + l_C 
\end{equation}
where between the flat adhered piece $l$ and a fully detached piece $l' = $QP, there is a circularly shaped curved piece $l_C = R \varphi$, 
with $R$ the radius of curvature and $\varphi$ the bending angle.  Reaching out to the suspension point P
the suspended piece is assumed to be straight, with the same slope as the tangent to the final circle point Q 
(see Fig. \ref{fig:elastic_mod}). That tangency condition determines $x(h)$:
\begin{equation}
\label{eq:tangence}
x(h)= R \sin\varphi + l' \cos\varphi
\end{equation}
Under the additional assumption that the flat-lying part $l$  enjoys the adhesion energy -$\epsilon$ per unit area, that is
-$\epsilon w$ per unit length, the total energy $E$ of the strip is a function of the lifting height $h$, of the adhered length $l$ and 
of the radius of the curved piece $R$ with bending angle $\varphi$:
\begin{equation}
E(l, R, \varphi) = - \epsilon w l  +  \frac{Kw\,\sin\varphi}{2R} 
\end{equation}
For a hard GNR ribbon that can be treated as approximately inextensible ~\cite{Lee08, Lopez15, Akinwande17}., the length constraint of eq. (\ref{eq:lengths}) must be added, 
yielding the ribbon enthalpy
\begin{equation}
F(h; l , R, \varphi, \mu) = E(l, R, \varphi) - \mu [L \, - \, l - R \varphi - l']                     
\end{equation}
where $l' =  [h - R(1 - \cos\varphi)]/\sin\varphi$ (see caption of Fig.3) and $\mu$ is a Lagrange multiplier. 
At this point, the ribbon equilibrium for a fixed height $h$ corresponds to the configurations $(l, R, \varphi)$ where 
Euler's equations
  \begin{align}
  \begin{split}
  \label{eq:Eulero}
    \frac{\partial F}{\partial l} &= \mu - \epsilon w= 0 \\        
    \frac{\partial F}{\partial R} &= -\frac{Kw \sin\varphi}{2R^2} - \mu \left(\frac{1 - \cos\varphi}{\sin\varphi} - \varphi \right) = 0 \\
    \frac{\partial F}{\partial \varphi} &= \frac{Kw \cos\varphi}{2R} + \mu \left(R + 
					   \frac{R(\cos\varphi - 1) - h\cos\varphi}{\sin^2\varphi} \right) = 0 \\
    \frac{\partial F}{\partial \mu} &= - \left(L \, - \, l\, - R \varphi - \frac{h - R(1 - \cos\varphi)}{\sin\varphi} \right) = 0. \\
    \end{split}
  \end{align}
are satisfied. From here it is straightforward to obtain the radius $R$ and the length $l$ of the adhered section of the ribbon as a function 
of bending angle $\varphi(h)$:
\begin{align}
  \begin{split}
 \label{eq:solutions}
 \mu &= \epsilon w\\
  R &= \frac{R_0}{\sqrt{2}} \frac{\sin\varphi}{\sqrt{\varphi \sin\varphi + \cos\varphi - 1}} \\
  l &= L - R \varphi - \frac{h – R(1 - \cos\varphi)}{\sin\varphi} \\
  &E_{\text{min}} = - \epsilon w \, l + \frac{Kw \sin\varphi}{2R}
 \end{split}
  \end{align}
where $R_0 = \sqrt{\frac{K}{\epsilon}}$ represents the value of the curvature radius for which $\varphi \rightarrow 0$, \textit{i.e.} in the 
case of an asymptotically flat ribbon, and $E_{\text{min}}$ the total energy at equilibrium. We note that $R_0$ depends on the ratio between the 
bending elasticity of the ribbon and the adhesion strength with the substrate, a feature repeatedly noted in recent MD simulations 
of of armchair graphene nanoribbons (GNRs) on a Au(111) substrate ~\cite{Gigli17, Gigli18}.  
For the ribbons used in experiment ~\cite{Kawai16}, the total bending elasticity of graphene is $K \sim 1.2$\,eV, the ribbon width is $w = 0.7$\,nm, and 
the adhesion $\epsilon w \sim 1.2$\,eV/nm. This yields a value of $R_0 = 0.84\,$nm, which is used to obtain the curves $\gamma (h)$ in Fig. 
~\ref{fig:elastic_mod}). These curves are in remarkable agreement with the GNRs equilibrium configurations at fixed height 
obtained in the realistic simulations of Refs. ~\cite{Gigli18, Gigli19}.
Substituting eq. (\ref{eq:solutions}) in the equation for the bending angle, one obtains the self-consistency equation for the angle $\varphi(h)$
\begin{equation}
 \label{eq:self_cons_phi}
 \frac{R_0}{2R} \cos\varphi + \left(\frac{R}{R_0} + \frac{\frac{R}{R_0} \left(\cos\varphi - 1 \right) - \frac{h}{R_0} 
 \cos\varphi}{\sin^2\varphi} \right) = 0
\end{equation}
where $R$  must satisfy eqs. (\ref{eq:solutions}). 
Eq. (\ref{eq:self_cons_phi}) can be solved numerically for the bending angle $\varphi$ for a grid of points of lifting heights $h$. 
In the case of graphene ribbons, the height ranges in the interval $0 \le h \le L$, where $L = 30.2$\,nm. 
We thus obtain a full numerical evaluation for all quantities of eqs. (\ref{eq:solutions}) for this specific case. 
We note, more generally, that eq. (\ref{eq:self_cons_phi}) only depends on the ratio $h/R_0$ and not separately on the two quantities. 
This means that the solution for the bending angle can be expressed generally as 
\begin{equation}
\label{eq:universal}
\varphi(h) = \Tilde{\varphi} \left(\frac{h}{R_0} \right)
\end{equation}
where $\Tilde{\varphi}$ is a universal function, independent of bending rigidity $K$ and of adhesive energy per unit area $\epsilon$. 
The functional form of our solutions is thus representative for a generic superlubric ribbon beyond the specific GNR case.

\section{Results and discussion}

%Fig. 4
\begin{figure}[!t]
\centering
\includegraphics[width = 0.50\textwidth]{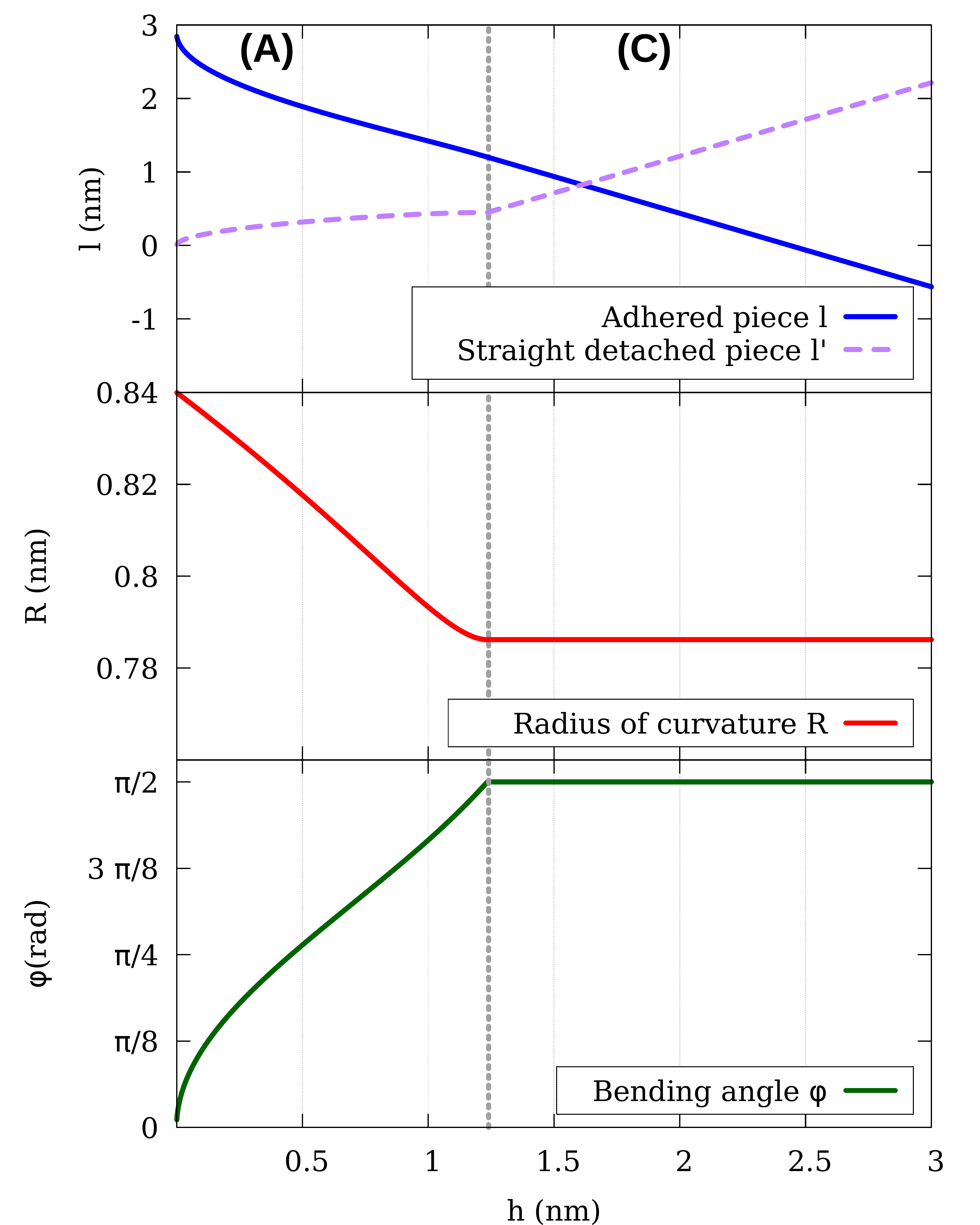}
\caption{
  Solution of eqs. (\ref{eq:solutions}) and (\ref{eq:self_cons_phi}). Two different regimes are visible: the initial prying (\text{A}) 
  in the range $0 < h \le h_C$ (where $h_C = 1.24$\,nm) marked by a rising of bending angle and a decreasing curvature radius;
  the steady peeling regime (\text{C}) for $h > h_C$. The transition between (\text{A}) and (\text{C}) is marked by a non-analytical kink 
  of the bending angle $\varphi(h)$ approaching $\pi/2$ and of the radius of curvature $R(h)$, both with a cusp at the critical height $h_C$.  
  The lengths of the adhered piece $l(h)$ and of the straight detached piece $l'(h)$, suddenly change slope at $h_C$. 
  Note that the curvature radius $R$ starts off at $R=R_0$, and not at infinity as one might have expected.
}
\label{fig:elastic_results}
\end{figure}
%

%Fig. 5
\begin{figure}[!t]
\centering
\includegraphics[width = 0.50\textwidth]{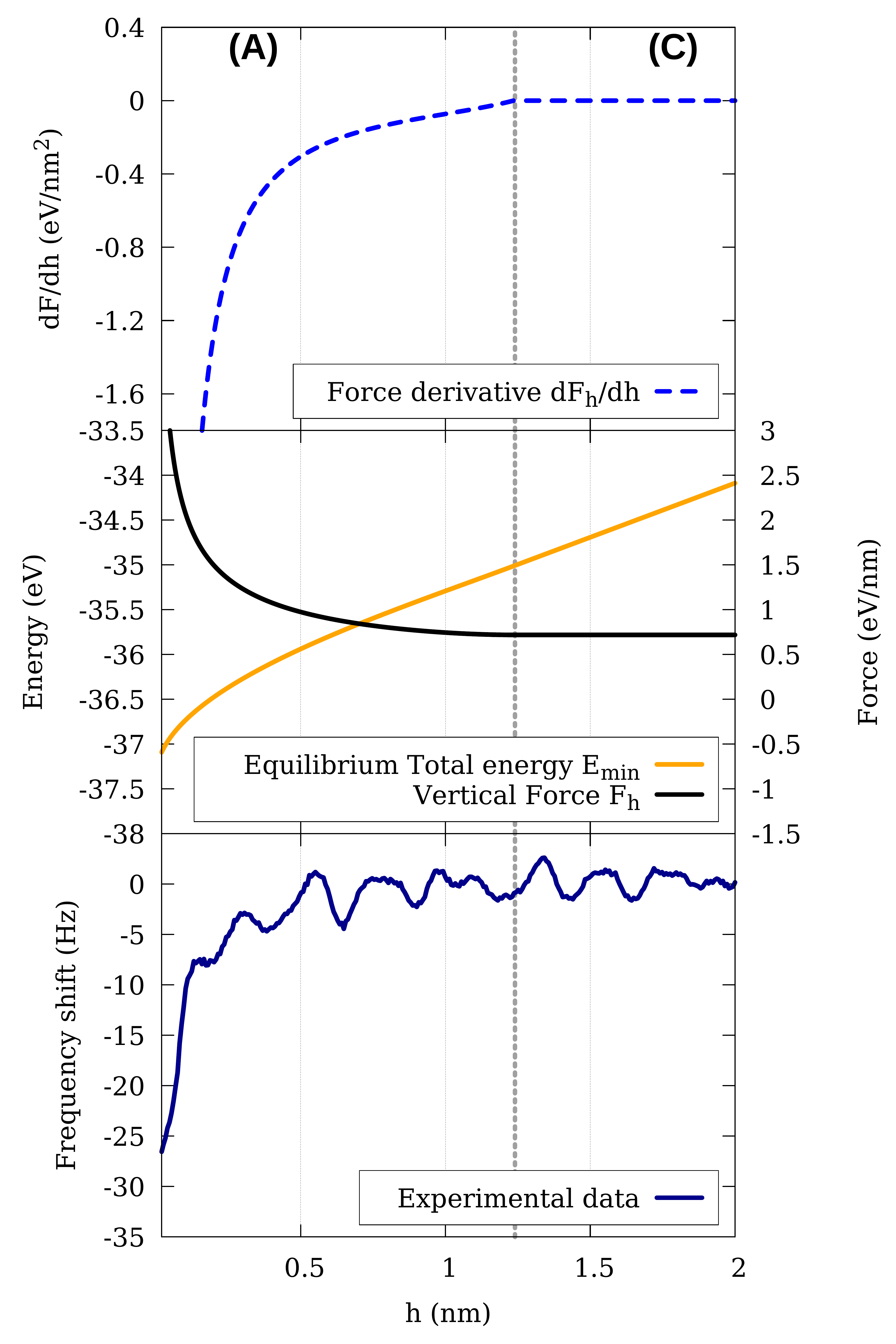}
\caption{
  Solution of eqs. (\ref{eq:solutions}) and (\ref{eq:self_cons_phi}) for the total potential energy, its derivative, \textit{i.e.}
  the vertical force upon detachment, and the force derivative $dF_h/dh$. The transition between (\text{A}) and (\text{C}) appears
  smoother than that of  angle and length . Yet, the vertical force derivative, directly comparable with the experimental  AFM frequency shift
  is non-analytic for $h \rightarrow h_C$. Bottom panel: experimental frequency shift for AFM lifting of a $\sim 30$ nm long GNR on Au(111), 
  from Ref. ~\cite{Gigli18}.
  The initial steep increase corresponds to the prying regime (\text{A}); the subsequent plateau is characteristic of regime (\text{C}). 
  The weak oscillations in the plateau reflect weak atomistic surface corrugations, here also involving the herringbone reconstruction of Au(111), 
  all neglected in our smooth model.
}
\label{fig:energy_vs_force}
\end{figure}

The results of eqs. (\ref{eq:self_cons_phi}) are shown in Fig. (\ref{fig:elastic_mod}), which exhibits the two different regimes. In the initial 
prying phase (\text{A}) the ribbon's curvature is building up from the initial $1/R_0$ as the bending angle $\varphi(h)$ grows. 
This regime ends at a critical height, $h_C = 1.24$\,nm with present parameters, where $\varphi(h)$ reaches $\pi/2$, 
and the detached ribbon is vertical. Inside phase (\text{A}), for $h < h_C$, the straight detached section 
GNR length does not depart appreciably from zero.  
For $h > h_C$ peeling enters regime (\text{C}) through a transition which in this model is of first order. 
Beyond that, the curvature radius $R(h)$ and the bending angle $\varphi(h)$ remain constant; in fact, the overall shape of the ribbon remains unaltered
during detachment. The adhered piece slides forward, consumed as it feeds into the $\varphi = \pi/2$ arc, thus steadily converted into straight 
lifted GNR, whose length increases linearly with $h$. The  (\text{A})-(\text{C}) transition can be identified in the form of crossover in 
our earlier atomistic simulations ~\cite{Gigli19}  (Fig. S2,  Supporting Information).
The agreement between experiment and our predicted transition scenario, including the critical height $h_C = 1.24\,$nm is remarkable.
As a general point, we note that the enthalpy minimization always has two solutions, one with $\varphi = \pi/2$  and another one for $\varphi = \varphi(h)$. 
They cross at $h=h_C$, so that below $h_C$ the latter prevails (phase (\text{A})), above $h_C$ the former (phase (\text{C})). 
In order to obtain a closer comparison with the experimental data of \cite{Gigli19}, in Fig. (\ref{fig:energy_vs_force}) we show the behavior of 
the total potential energy at equilibrium $E_{\text{min}}(h)$, the vertical force upon detachment predicted by the model $F_h = dE_{\text{min}}/dh$ and 
finally the vertical force gradient $dF_h/dh = d^2E_{\text{min}}/dh^2$, a quantity proportional to the measured AFM frequency shift.
The total potential energy of the ribbon increases linearly with $h$ in the sliding phase (\text{C}), consistent with the vertical 
pulling force, approaching steady value $F_{\infty} = 0.5$\, eV/nm. The non-analytic kink at the (\text{A})-(\text{C}) transition where the 
bending angle hits $\pi/2$ is here reflected in the vertical force gradient for $h \rightarrow h_C$. The same singularity affects the second derivative 
of the effective adhesive length $l$. The presence of corrugation and other sources of noise in AFM peeling data turns the relatively weak 
force singularity into a crossover.  

\section{Conclusions}

We present and analytically solve a model describing the peeling from a flat substrate of narrow, flexible, 
structurally lubric nanoribbons and nanostructures, inspired by experimental and simulation evidences 
obtained from graphene nanoribbons on gold surfaces ~\cite{Kawai16, Gigli17, Gigli19, Pawlak16}. 
The model predicts two regimes, denoted as (\text{A}) and (\text{C}), in Fig. (\ref{fig:elastic_mod}) and 
Fig. (\ref{fig:energy_vs_force}), separated by a first-order phase transition. 
If the equilibrium configuration of the ribbon is supposed to evolve adiabatically for increasing height $h$ 
as one extreme is lifted up, for instance by the vertical retraction of an AFM tip, then (\text{A}) represents 
the initial phase in which the vertical force of the AFM is converted into mechanical deformations of the bent section,
with a steady increase of the bending angle, 
and a small forward sliding of the adhered part, as required by inextensiblility. 
%, and no sliding of the adhered part. 
%
At the critical height $h_C$, the process enters regime (\text{C}), where the bending angle saturates at $\pi/2$ and 
the adhered part slides forward as $h$ increases. Both features have been repeatedly observed both in experiments 
and in simulations. 
The singularity that separates the two regimes is in this model of first order, as reflected in all physical quantities: 
the radius of curvature $R$,the bending angle, the lifting force derivative.  
Instructive, analytical, and relatively general as it is, the model of course presents important limitations. First and foremost, 
it neglects corrugations, that by causing atomistic interlocking may lead to replace regime (\text{C}) by a stick-slip alternation 
of regimes  (\text{B}) - (\text{C}), as observed in simulations ~\cite{Gigli19}. 
On the other hand, as explained in section \ref{sec:Intro}, phase (\text{B}),  where peeling proceeds by stripping 
%and
without 
sliding, is a behavior that does not occur on a totally flat substrate. A second limitation is that the bent section may deviate from 
circularity and the detached piece may not be as straight as assumed. Removing these geometrical limitations could in principle
smoothen out the first order (\text{A})-(\text{C}) singularity into a crossover. 
In spite of all that, the  striking simplicity and universal analytical predictions will make this model an important starting point 
for the understanding of state-of-the-art detachment of generic nanoribbons and other lubric nanostructures, and a necessary 
alternative to Kendall's model for the opposite, strongly corrugated and adhesive limit.

\vspace*{0.5 cm}

\section*{Acknowledgments}

\noindent We acknowledge funding from the European Research Council (ERC) under FP7 Advanced
Grant No. 320796 MODPHYSFRICT and  under Horizon 2020 Advanced Grant No. 824402 ULTRADISS.
\noindent Informative and collaborative discussions are gratefully acknowledged with E. Meyer, S. Kawai, and R. Pawlak, and 
with R. Guerra and N. Manini.

\vspace{3mm}

\newcommand*{\xdash}[1][3em]{\rule[0.5ex]{#1}{0.55pt}}
\xdash[10em]

\end{document}